\documentclass[prd,aps,eqsecnum,floatfix,nofootinbib,preprint,tightenlines]{revtex4}

\usepackage{latexsym}
\usepackage{amsmath}
\usepackage{hyperref}  

\def\bibi{\bibitem}

\def\a{\alpha}

\def\c{\chi}
\def\d{\delta}
\def\e{\epsilon}                
\def\g{\gamma}

\def\j{\psi}

\def\l{\lambda}
\def\m{\mu}
\def\n{\nu}

\def\p{\pi}                     
\def\t{\tau}

\def\x{\xi}

\def\L{\Lambda}

\def\P{\Pi}

\def\S{\Sigma}

\def\cu{{\cal U}}
\def\cv{{\cal V}}

\def\cbo{{\,\raise-.15ex\Sc [\,}}                       

\def\svev#1{\left\langle #1\right\rangle}       

\def\ddt#1{{\buildrel {\hbox{\LARGE .\kern-2pt.}} \over {#1}}}

\def\eg{\mbox{\it e.g.}}

\def\tr{{\rm tr}\,}

\def\hc{{\rm h.c.\,}}

\def\half{{1\over 2}}

\def\ttl#1{{\it #1}}

\long\def\symbolfootnote[#1]#2{\begingroup%
\def\thefootnote{\fnsymbol{footnote}}\footnote[#1]{#2}\endgroup}

\long \def \blockcomment #1\endcomment{}

\def\seef{{\it cf.\  }}

\def\bc{\overline{\c}}
\def\bj{\overline{\j}}
\def\bl{\overline{\l}}

\def\tP{\tilde{P}}
\def\tT{\tilde{T}}

\begin{document}

\begin{center}
\vspace{10mm}
{\large\bf
Vacuum alignment and lattice artifacts: staggered fermions
}
\\[12mm]
Maarten Golterman$^a$
and Yigal Shamir$^b$
\\[8mm]
{\small\it
$^a$Department of Physics and Astronomy\\
San Francisco State University, San Francisco, CA 94132, USA}%
\\[5mm]
{\small\it $^b$Raymond and Beverly Sackler School of Physics and Astronomy\\
Tel Aviv University, Ramat~Aviv,~69978~ISRAEL}%
\\[10mm]
{ABSTRACT}
\\[2mm]
\end{center}

\begin{quotation}
In confining lattice gauge theories in which part of the flavor group
is coupled weakly to additional gauge fields, both the dynamics of the
weak gauge fields as well as lattice artifacts may have non-trivial
effects on the orientation of the vacuum in flavor space.   Here we
discuss this issue for lattice gauge theories employing staggered
fermions.   Staggered fermions break flavor symmetries to a much
smaller group on the lattice, and orientations in flavor space that are
equivalent in the continuum may be distinct on the lattice.
Assuming universality, we show that in the continuum limit the weakly gauged
flavor symmetries are always vector-like, disproving a recent claim
in the literature.
\end{quotation}

\renewcommand{\thefootnote}{\arabic{footnote}} \setcounter{footnote}{0}

\newpage
\section{\label{Intro} Introduction}
Recently, there has been a growing interest in the non-perturbative
study of gauge theories with both strong and weak gauge interactions---with the
latter being weak at the scale where the former is strong---encompassing
both Standard-Model and
beyond-the-Standard-Model physics.   An example of the first is the
inclusion of electromagnetic effects in lattice QCD \cite{NTlat13}, and an example
of the second is the study of composite Higgs models \cite{MP2005,RC2010},
where a relatively light Higgs particle
is assumed to arise as a Nambu--Goldstone boson (NGB) of some new strong
dynamics.  In such models, some of the flavor symmetries of the new strong
sector are weakly coupled to additional gauge fields, turning
the Higgs particle into a pseudo-NGB, which then induces
electro-weak symmetry breaking dynamically.

On the lattice, most discretizations of the fermion action
typically have a much reduced flavor symmetry in comparison to the corresponding
continuum theory.  However, for commonly used fermion discretizations
the full continuum symmetry gets restored in the continuum limit with
minimal or no fine tuning. Indeed a large variety of lattice fermion actions
gives rise to the same
theory in the continuum limit, a phenomenon known as universality.

In order to make the discussion more concrete, we will limit
ourselves here to confining
$SU(N_c)$ gauge theories with $N_c\ge 3$, coupled to an
even number $N_f$ of Dirac fermions in the fundamental representation.\footnote{We anticipate that the generalization to other groups and
representations is relatively straightforward in many cases.}
Any such theory can be formulated on the lattice using
reduced staggered fermions \cite{STW1981,KS1981,KMNP1983,DS1983,GS1984}.
If $N_f$ is a multiple of 4,
standard staggered fermions can be used.\footnote{%
  Large-scale numerical simulations of QCD make use of 3
  standard staggered fields (one for each of the up, down and strange quarks),
  but the fourth root of the fermion determinant is taken in order
  to reduce the number of fermion species from 12 to 3
  (see, \eg, Refs.~\cite{rooted,stagrev} and references therein).
  In this paper we only consider local lattice theories, avoiding
  any fractional powers of the staggered-fermion determinant.
}
In the massless limit, the flavor symmetry of the continuum theory is
$SU(N_f)_L\times SU(N_f)_R$.  The use of reduced or standard
staggered fermions leaves intact only a rather small subgroup of the flavor
symmetry, but the remaining flavor symmetries are automatically restored
in the continuum limit \cite{DS1983,GS1984}.

There is a less well known aspect of staggered fermions.
The physical role of the continuous global symmetries of the massless
lattice theory, or lattice flavor symmetries for short,
depends on the choice of lattice mass terms.
The unit cell of the staggered action is a $2^4$ hypercube,
and mass terms coupling any two lattice sites within the unit cell
can be written down \cite{DS1983,GS1984}.  For the most common,
same-site mass term, some of the lattice flavor symmetries become
axial symmetries in the continuum limit.  However, for mass terms
that couple pairs of lattice sites separated by an odd number of links,
all the lattice flavor symmetries become vector symmetries
in the continuum limit.  Thus, even if all fermions have equal masses,
the embedding of the lattice flavor symmetries in
the continuum flavor group $SU(N_f)_L\times SU(N_f)_R$ depends on
the choice of lattice mass terms.

So far, these observations were essentially just technical.  The situation changes
if a subgroup of the continuum flavor symmetry is weakly coupled
to new dynamical gauge fields (``weak gauge fields,'' for short).
The new dynamics will typically distinguish between different orientations
of the mass terms, or, in the massless limit, of the fermion condensate.
Orientations of the fermion condensate that do not break spontaneously
any of the weakly gauged symmetries will be energetically favorable,
a phenomenon known as vacuum alignment \cite{MP1980}.
Away from the chiral limit, there will be competing effects
between the explicit mass terms on the one hand,
and the effective potential induced by the weak gauge fields on
the other hand.  The outcome---the orientation of the vacuum---will depend
on the details.

A third source of dynamical effects is provided by the discretization itself.
The reduced symmetry of the lattice theory allows for the dynamical
generation of an  effective potential at order $a^2$, where $a$ is the
lattice spacing.  This effective potential, too, can give rise to a non-trivial
phase diagram.  The most familiar example of this sort is the so-called
Aoki phase encountered for Wilson fermions \cite{Aoki1983,Creutz1995,ShSi1998},
where some of the lattice (vector) flavor symmetries undergo
spontaneous breaking.  An order-$a^2$ effective potential gets
generated for staggered fermions as well \cite{LS1999,AB2003}, leading
to the possibility of similar phases \cite{AW2004}.

When studying composite-Higgs models, or any other model involving
the dynamical breaking of electro-weak symmetry, we have to take
the combined continuum and chiral limit.  The relevant
phase diagram is therefore controlled by the dynamics of the weak gauge fields
only.  However,  realistic lattice simulations are carried out
away from both limits.  In the lattice simulation, all three sources:
explicit mass terms, weak gauge fields, and discretization effects,
will in general compete, leading to a potentially very complicated outcome.
Both the continuum and the chiral limits will have to be studied
with great care, in order to determine whether we have arrived close
enough to the combined limit such that
the weak gauge field dynamics has taken over.

In this paper we study these questions using chiral lagrangian techniques
\cite{MG2009}.  After a brief review of relevant facts about
staggered fermions in Sec.~\ref{basics}, we turn in Sec.~\ref{8 flavors}
to the 8-flavor theory.  This theory can be formulated on the lattice
using two standard staggered fields, or, equivalently,
four reduced staggered fields.
Starting with the case that none of the flavor symmetries are gauged
we compare two different choices:
the same-site and the one-link mass terms.
While the continuum limit is the same for both choices,
only in the case of the same-site mass term do some of the lattice
flavor symmetries turn into axial symmetries of the continuum theory.

We then study what happens when the lattice flavor symmetries are weakly
gauged.  Using Witten's inequality \cite{EW1983}
we prove that, after taking the continuum and chiral limits,
the vacuum state orients itself along the one-link mass term.
Therefore all of the weakly gauged symmetries are vectorial, and none of them
are broken spontaneously,
in agreement with the Vafa-Witten theorem \cite{VW1983}.
This result refutes a claim recently made in the literature \cite{CV2013}.
In Sec.~\ref{6 flavors} we study the 6-flavor theory, with the new element
that in this case the reduced staggered formalism is indispensable, and
we arrive at
similar conclusions.  We conclude in Sec.~\ref{conclusion}.  In App.~\ref{Veffcont}
we rederive the continuum effective potential, while App.~\ref{Veffasq}
contains some simple observations which follow from the structure of
the order-$a^2$ effective potential for the 8-flavor theory.

\section{\label{basics} Staggered-fermions basics}
In this section, we review some of the basic properties of staggered
fermions.   For a comprehensive treatment, we refer to Refs.~\cite{DS1983,GS1984},
and to the reviews in Refs.~\cite{stagrev,MG2009}.

The lagrangian for a single massless staggered fermion $\c(x)$ coupled to a
gauge field $U_\m(x)$ is
\begin{equation}
\label{Sstag}
S=\half\sum_{x\m}\eta_\m(x)\,\bc(x)\left(U_\m(x)\c(x+\m)-
U^\dagger_\m(x-\m)\c(x-\m)\right)\ ,
\end{equation}
in which the phase factors
\begin{equation}
\label{eta}
\eta_\m(x)=(-1)^{x_1+\dots+x_{\m-1}}\ ,  \qquad \m=1,\ldots,4 \ ,
\end{equation}
take over the role of the Dirac matrices.  Along with a suitable pure-gauge
action, the staggered-fermion action~(\ref{Sstag})
gives rise to a gauge theory with
four massless Dirac flavors all in the same representation of the gauge group
in the continuum limit.\footnote{In the context of QCD, usually these
four flavors are referred to as ``tastes,'' but here we will choose to
refer to them as flavors.}   The continuum theory thus has an
$SU(4)_L\times SU(4)_R$ flavor symmetry.

Apart from fermion number, the lattice action~(\ref{Sstag})
has only one continuous symmetry, $U(1)_\e$, given by \cite{KS1981}
\begin{equation}
\label{U1eps}
\c(x)\to e^{i\a\e(x)}\c(x)\ ,\qquad\bc(x)\to\bc(x)e^{i\a\e(x)}\ ,
\end{equation}
with
\begin{equation}
\label{eps}
\e(x)=(-1)^{x_1+x_2+x_3+x_4}\ .
\end{equation}
This symmetry is usually interpreted as an axial symmetry, but this
interpretation actually depends on the mass terms that are added
to the lattice theory.  In most applications,
a single-site mass term $m\bc(x)\c(x)$ is chosen.
This breaks $U(1)_\e$ softly, signifying that $U(1)_\e$ is indeed
an axial symmetry in this case.

However, one may choose different mass terms.
For instance, another gauge-invariant mass term is given by
\begin{equation}
\label{onelinkmass}
S_{\rm 1-link}=\half\sum_{x\m}m_\m\zeta_\m(x)\,\bc(x)\left(U_\m(x)\c(x+\m)+
U^\dagger_\m(x-\m)\c(x-\m)\right)\ ,
\end{equation}
with a new set of phase factors
\begin{equation}
\label{zeta}
\zeta_\m(x)=(-1)^{x_{\m+1}+\dots+x_{4}}\ .
\end{equation}
These phase factors ensure that $S_{\rm 1-link}$ is
invariant under hypercubic rotations if $m_\mu$ is treated
as a vector spurion.  Since $S_{\rm 1-link}$ couples fermion and
anti-fermion fields
that are one link apart, it is invariant under $U(1)_\e$, which implies that
the $U(1)_\e$ symmetry ends up as a
vector symmetry in the continuum limit \cite{DS1983,GS1984}.
In this limit the four flavors are degenerate, and their
common mass is proportional to $m_{\rm 1-link}=\sqrt{\sum_\m m_\m^2}$.
The one-link mass term~(\ref{onelinkmass})
is particularly relevant in the case of reduced staggered fermions,
which we introduce next.

Let us project the field $\c(x)$ onto the even sites, and
the independent field $\bc(x)$ onto the odd sites:
\begin{equation}
\label{proj}
\c^+(x) = \half(1+\e(x))\c(x)\ ,\qquad\bc^-(x)=\half(1-\e(x))\bc(x)\ ,
\end{equation}
thereby thinning out the number of degrees of freedom by a factor two.
Applying this projection to the action~(\ref{Sstag})
gives rise to the (massless) reduced staggered fermion action
\begin{equation}
\label{Sred}
S^+=\half\sum_{x\m}\eta_\m(x)\,\bc^-(x)\left(U_\m(x)\c^+(x+\m)-
U^\dagger_\m(x-\m)\c^+(x-\m)\right)\ .
\end{equation}
Instead of four, this action gives rise to two Dirac flavors in the
continuum limit, with flavor symmetry group $SU(2)_L\times SU(2)_R$
\cite{STW1981,DS1983,GS1984}.

A different reduced staggered action is obtained by
reversing the projections in Eqs.~(\ref{proj}) and~(\ref{Sred}), namely, by choosing
\begin{equation}
\label{opproj}
\c^-(x) = \half(1-\e(x))\c(x)\ ,\qquad\bc^+(x)=\half(1+\e(x))\bc(x)\ .
\end{equation}
We may take two reduced staggered fields, one of each type, and
re-assemble them into a single standard staggered fermion.
The same-site mass term we have discussed for the standard case
decomposes as
\begin{equation}
\label{ssmasseq}
m\left[\bc^+(x)\c^+(x)+\bc^-(x)\c^-(x)\right]=m\bc(x)\c(x)\ ,
\end{equation}
showing that the two reduced-staggered types
defined by the projections~(\ref{proj}) and~(\ref{opproj})
are coupled to each other.
In contrast, the one-link mass term of Eq.~(\ref{onelinkmass}) involves
no coupling between the two reduced staggered types.

Let us elaborate on this observation.
Given a single reduced staggered field, it is evidently
not possible to construct a same-site mass term.
The simplest mass term is the one-link mass term obtained
from Eq.~(\ref{onelinkmass}) above via the relevant projection
\begin{equation}
\label{Sredmass}
S_{\rm 1-link}^\pm = \half\sum_{x\m}m_\m\zeta_\m(x)\,
\bc^\mp(x)\left(U_\m(x)\c^\pm(x+\m)+U^\dagger_\m(x-\m)\c^\pm(x-\m)\right)\ .
\end{equation}
An independent mass term can be constructed by coupling
fermion and anti-fermion fields that are three links apart.\footnote{%
  Replacing the one-link mass terms by three-link mass terms
does not change our conclusions.  We will therefore limit the discussion
to one-link mass terms.
}
Either way, the number of links separating the reduced fermion and
anti-fermion fields has to be odd,
and therefore any mass term in the reduced case is invariant under $U(1)_\e$.

It follows that whenever the lattice theory does not involve
bilinear couplings between reduced staggered fields of different types,
all the lattice flavor symmetries necessarily turn into vector symmetries
in the continuum limit.  Indeed, considering
the standard staggered action~(\ref{Sstag}) let us denote
the generators of fermion number and of $U(1)_\e$ by $Q_s$
and $Q_\e$ respectively.  It is easily seen that the linear combinations
$Q_s+Q_\e$ and $Q_s-Q_\e$ generate the fermion number symmetries associated with
the projections~(\ref{proj}) and~(\ref{opproj}), respectively.
Provided that the chosen mass terms respect the individual
fermion number symmetries, these symmetries are, therefore, vectorial.

One can construct theories with an arbitrary even number of flavors, $N_f$,
using $N^+ \le N_f/2$ reduced staggered fields of type~(\ref{proj}),
together with $N^-=N_f/2-N^+$ reduced fields of type~(\ref{opproj}).
The same set of fields can also be regarded as consisting of
$N_s=\min(N^+,N^-)$ standard staggered fields, with the remaining
reduced staggered fields being all of the same type.
In the massless case, the lattice flavor symmetry is
$U(N^+)\times U(N^-)$, generalizing the fermion number symmetries
of the individual reduced fields.  The flavor symmetry remains intact if
one-link mass terms with the same vector $m_\m^\pm$  are introduced
for all reduced fields of a given type, consistent with the fact
that in this case, all the lattice flavor symmetries are vectorial.
For other choices of mass terms, some of the lattice flavor symmetries
may be softly broken.

As a final comment we note that, thanks to additional discrete symmetries,
the renormalization of all mass terms for both standard and reduced
staggered fields is multiplicative \cite{GS1984}.  The chiral limit
is therefore well-defined at non-zero lattice spacing, and corresponds
to the vanishing of all bare mass terms.

In the next two sections, we will employ these observations
in the context of the 8-flavor and 6-flavor theories.   The 8-flavor theory
we will consider
corresponds to the choice $N^+=N^-=2$, whereas the 6-flavor theory
corresponds to $N^+=2$, $N^-=1$.

\section{\label{8 flavors} Eight flavors}
In this section, we will consider an 8-flavor theory coupled to a strong
gauge field $U_\m(x)$.  The lattice theory is constructed using two
standard staggered fields, or, equivalently, four reduced staggered
fields, two of each type.  For clarity, we will denote
by $\c_i,\bc_i$, $i=1,2$, the two reduced staggered fields of type~(\ref{proj}),
and by $\l_i,\bl_i$, $i=1,2$, the two reduced fields of type~(\ref{opproj}).
According to the discussion in the previous section, if we disregard
$U(1)$ factors, the non-abelian global symmetry of the massless lattice
theory is $SU(2)_\c\times SU(2)_\l$.  In the continuum limit,
the flavor symmetry enlarges to $SU(8)_L\times SU(8)_R$,
which we will assume to be spontaneously broken
to the diagonal $SU(8)_V$ subgroup.

We will consider two choices for the lattice mass term,
as well as the corresponding
orientations of the fermion condensate in the chiral limit.  While both
choices give rise to the same continuum limit, the embedding
of the lattice flavor symmetries inside the continuum symmetry group
is different.  For one of these choices, some of the lattice flavor symmetries
become spontaneously broken axial symmetries in the continuum limit;
for the other choice, all the lattice flavor symmetries are
vectorial in the continuum limit.
We will then weakly couple all the (non-abelian)
lattice flavor symmetries to additional gauge fields, and prove that
in this case all of them become unbroken
vectorial symmetries of the continuum theory.

In the continuum limit, each reduced staggered field gives rise to
two Dirac fields, according to
\begin{equation}
\label{id}
\c_1\to\j_1,\ \j_2\ ,\quad \c_2\to\j_3,\ \j_4\ ,\quad \l_1\to\j_5,\ \j_6\ ,\quad\l_2\to\j_7,\ \j_8\ .
\end{equation}
Alternatively, viewing the fermion content as two standard staggered fields
$\c_i+\l_i$, the continuum flavors $\j_1$, $\j_2$, $\j_5$ and $\j_6$
emerge from $\c_1+\l_1$, while $\j_3$, $\j_4$, $\j_7$ and $\j_8$
emerge from $\c_2+\l_2$.

Our first choice for the mass terms is to use
the one-link mass term~(\ref{Sredmass}) for each
reduced staggered fermion, always with the same parameters $m_\m$.
In the continuum theory, the resulting mass term is
\begin{equation}
\label{onelinkcont}
  \sum_{i=1}^2 S_{\rm 1-link}(\c_i,\l_i;m_\m)
  \to m\int d^4x\sum_{k=1}^8\bj_k\j_k\ ,
\end{equation}
where $m\ge0$ is given by\footnote{%
  We disregard the (multiplicative) renormalization
of the mass parameters.
}
\begin{equation}
\label{mmu}
  m^2 = \sum_\m m_\m^2\ .
\end{equation}
If we arrange the continuum Dirac fields into a vector,
the continuum mass matrix is proportional to the identity matrix,
\begin{equation}
\label{olmatrix}
M_1=m\, I_8\ ,
\end{equation}
where $I_n$ denotes the $n\times n$ identity matrix.\footnote{%
  In the continuum limit, a basis can always be chosen for the
two Dirac fields originating from a given reduced staggered field
such that the mass matrix takes the form~(\ref{olmatrix})
by construction \cite{DS1983,GS1984}.
}

Alternatively, we can use the single-site mass term~(\ref{ssmasseq}), obtaining
\begin{eqnarray}
\label{sscont}
&&m\sum_x\left(\bl_1\c_1+\bl_2\c_2+\bc_1\l_1+\bc_2\l_2\right)\to\\
&&m\int d^4x\left(\bj_5\j_1+\bj_6\j_2+\bj_7\j_3+\bj_8\j_4+\bj_1\j_5+\bj_2\j_6+\bj_3\j_7+\bj_4\j_8
\right)\ .\nonumber
\end{eqnarray}
The corresponding mass matrix can be written in the form
\begin{equation}
  M_0 =m\, \t_1 \otimes I_2 \otimes I_2 =m\, \t_1 \otimes I_4 \ .
\label{massmat}
\end{equation}
In this notation, any $8\times 8$ matrix is expressed as a sum
of tensor products.  Each tensor product consists of three terms,
each of which can be one of the Pauli matrices $\t_a$, $a=1,2,3,$
or the identity matrix $I_2$.  The index of the first $2\times 2$ matrix
in the tensor product identifies the reduced staggered type ($\c$ or $\l$)
from which the continuum flavor originates,
and the associated projectors are
\begin{eqnarray}
\label{Ppm}
  P_\c &=& \half (I_2+\t_3) \otimes I_4 \ \equiv \ \tP_\c \otimes I_4 \ ,
\\
  P_\l &=& \half (I_2-\t_3) \otimes I_4 \ \equiv \ \tP_\l \otimes I_4 \ .
\nonumber
\end{eqnarray}
The index of the second factor in the tensor product is the flavor
index of the corresponding reduced staggered type, while the index
of the last factor runs over the two continuum flavors
that emerge from a given reduced staggered field.

In the continuum limit, the two choices for the mass term are equivalent.
Indeed, one can rotate $M_0$ to the standard form~(\ref{olmatrix})
by a non-anomalous transformation $U\in SU(8)_L\times SU(8)_R$, under which
\begin{equation}
\label{trans}
  \j \to U\j \ ,\qquad
  \bj \to \bj \g_0 U^\dagger \g_0 \ .
\end{equation}
To this end, we first apply the purely vectorial transformation
\begin{equation}
  P     = \frac{1}{\sqrt{2}}\,  (I_2-i\t_2) \otimes I_4 \ ,
\label{P}
\end{equation}
so that now
\begin{equation}
  P^\dagger M_0 P = m\ \t_3 \otimes I_4 =
  m\left(\begin{array}{cc}
    I_4 & 0 \\
    0  & -I_4
  \end{array}\right)\ .
\label{M2}
\end{equation}
In order to rotate the lower-right block from $-I_4$ into $+I_4$
we apply the non-anomalous axial rotation
\begin{equation}
\label{Q}
  Q = P_\c + i\g_5 \tP_\l \otimes\t_3\otimes I_2 =
  \left(\begin{array}{cc}
    I_4 & 0 \\
    0  & i\g_5 \t_3 \otimes I_2
  \end{array}\right)\ .
\end{equation}
Using Eq.~(\ref{trans}) we arrive at
\begin{equation}
   QP^\dagger M_0 P Q =m\, I_8 \ ,
\label{finalM}
\end{equation}
thereby reproducing Eq.~(\ref{olmatrix}), but now for the same-site mass term.

\subsection{\label{flavor} Global lattice flavor symmetry}
Let us now discuss the interplay of the $SU(2)_\c\times SU(2)_\l$
lattice flavor symmetry and the two mass terms.
We are interested in the fate of these symmetries after
taking the continuum limit,
followed by the chiral limit where
the mass term is turned off (after the infinite-volume limit has been taken).

The one-link mass term~(\ref{onelinkcont}) respects the full lattice
flavor symmetry.  On the basis of continuum fields introduced in Eq.~(\ref{id}),
the resulting mass matrix~(\ref{olmatrix})
is proportional to the identity matrix,
whereas the $SU(2)_\c\times SU(2)_\l$ generators take the form
\begin{subequations}
\label{su2s}
\begin{eqnarray}
  T_a^\c &=& \tP_\c \otimes \t_a\otimes I_2 \ ,
\label{su2sa}\\
  T_a^\l &=& \tP_\l \otimes \t_a\otimes I_2 \ .
\label{su2sb}
\end{eqnarray}
\end{subequations}
All six generators are vectorial in this case,
as they are proportional to the identity matrix in Dirac space.

We next turn to the same-site mass term~(\ref{sscont}).
The diagonal subgroup generated by $T_a^\c + T_a^\l$
commutes
with this mass term.  The other three generators, $T_a^\c - T_a^\l$,
are proportional to the phase factor $\e(x)$ of Eq.~(\ref{eps}),
and are broken by the same-site
mass term.  On the same continuum basis, these
linear combinations take the form
\begin{subequations}
\label{av}
\begin{eqnarray}
  T_a^+ &=& T_a^\c + T_a^\l \ = \ I_2 \otimes \t_a \otimes I_2 \ ,
\label{ava}\\
  T_a^- &=& T_a^\c - T_a^\l \ = \ \t_3 \otimes \t_a \otimes I_2 \ .
\label{avb}
\end{eqnarray}
\end{subequations}
As on the lattice, the $T^+_a$
commute with the mass matrix~(\ref{massmat}), whereas the $T^-_a$
do not.  Applying the basis transformation that brings the
mass matrix~(\ref{massmat}) to the diagonal form~(\ref{finalM}),
the generators become
\begin{subequations}
\label{PQva}
\begin{eqnarray}
  Q^\dagger P^\dagger T_a^+ P Q &=&
  \Big( \tP_\c \otimes \t_a + \tP_\l \otimes \t'_a \Big) \otimes I_2 \ ,
\label{PQv}\\
  Q^\dagger P^\dagger T_a^- P Q &=&
  \g_5 \Big( \e_{ab3}\, \t_1 \otimes \t_b - \d_{a3} \t_2 \otimes I_2
  \Big) \otimes I_2\ ,
\label{PQa}
\end{eqnarray}
\end{subequations}
where $\t'_a = \t_3 \t_a \t_3$.  While the $T_a^+$ still generate
a vectorial symmetry, the $T_a^-$ now generate an axial symmetry.

The fact that we can rotate the mass matrix~(\ref{massmat}) to the diagonal
form~(\ref{finalM}) using an $SU(8)_L\times SU(8)_R$ transformation
implies that the two mass matrices are equivalent.
So are the corresponding orientations of the fermion condensate in
the chiral limit.
Indeed, with the restriction to a degenerate mass for all 8 flavors,
all possible choices for the lattice mass terms are equivalent in that,
in the continuum limit, the resulting symmetry-breaking
pattern is always $SU(8)_L\times SU(8)_R\to SU(8)_V$, with
the unbroken $SU(8)_V$ commuting with the mass matrix.
Any violation of this observation would constitute a violation of
universality.

But the fate of the lattice flavor symmetries is not the same.
In the case of the one-link mass term~(\ref{onelinkcont}), all of them
become unbroken vectorial symmetries of the continuum theory, whereas in
the case of the same-site mass term~(\ref{massmat}), this is true
only for half of the lattice symmetries, while the other half turn
into axial symmetries, which are spontaneously broken in the chiral limit.

\subsection{\label{gauge} Gauging $SU(2)_\c\times SU(2)_\l$}
We now introduce a new element, by promoting the lattice global symmetry group $SU(2)_\c\times SU(2)_\l$
to a local symmetry.  We introduce a dynamical gauge field $V_{\m a}$ minimally coupled
to the conserved currents of $SU(2)_\c$ with coupling constant $g_\c$,
and, similarly, a gauge field $W_{\m a}$ with coupling $g_\l$ for $SU(2)_\l$.
We will assume that both of the new couplings are weak at the scale $\L$ where
the original strong dynamics of the gauge field $U_\m$ is confining.\footnote{%
  For $N_c=3$ it is not clear whether or not the 8-flavor theory is confining
\cite{Nf8}.  The 8-flavor theory confines in the large $N_c$ limit,
and we will assume that $N_c$ is large enough that this is the case.
}

The effective low-energy theory depends on a non-linear field $\S(x)\in SU(8)$.
We may think of $\S_{k\ell}(x)$ as representing
the composite operator $\tr((1-\g_5)\j_k(x)\bj_\ell(x))$,
where the trace is over Dirac and strong gauge group indices.
The interaction with the dynamical weak gauge fields induces an
effective potential for the continuum theory.
To lowest non-trivial order in the weak gauge couplings,
the effective potential is \cite{MP1980}
\begin{equation}
\label{Vweak}
V_{\rm weak}(\S)=-g_\c^2 C\,\sum_a\tr\left(\S T^\c_a\S^\dagger  T^\c_a\right)
-g_\l^2 C\,\sum_a\tr\left(\S T^\l_a\S^\dagger  T^\l_a\right)\ .
\end{equation}
We have restricted the non-linear field to a constant value $\S(x)=\S$
representing the vacuum.  In the chiral limit of the continuum theory
there are no other effects of similar magnitude, and the vacuum state
is determined by minimizing $V_{\rm weak}(\S)$.

Let us now compare the vacua $\S_0$, defined by the orientation of
the same-site mass matrix~(\ref{massmat}), and $\S_1$,
defined by the orientation of the one-link mass matrix~(\ref{olmatrix}).
On the continuum basis introduced in Eq.~(\ref{id}), these vacua are
\begin{equation}
\label{candvac8}
\S_0=\t_1\otimes I_4\ ,\qquad \S_1=I_8\ .
\end{equation}
We find that
\begin{subequations}
\label{Vweakvalues}
\begin{eqnarray}
V_{\rm weak}(\S_0)&=&0\ ,
\label{Vweakvaluesa}\\
V_{\rm weak}(\S_1)&=&-12\left(g_\c^2+g_\l^2\right)C\ .
\label{Vweakvaluesb}
\end{eqnarray}
\end{subequations}
A key observation is that the low-energy constant $C$ is positive \cite{EW1983}.
The $\S_1$ vacuum wins, and, in fact, since Eq.~(\ref{Vweakvaluesb}) is the
minimum value $V_{\rm weak}$ can take, $\S_1$ is the correct vacuum
of the continuum theory.  The full vacuum manifold consists of all
$\S\in SU(8)$ where $V_{\rm weak}$ retains the value~(\ref{Vweakvaluesb}),
and therefore any representative of the true vacuum must commute
with all $T^{\c,\l}_a$.

In accordance with Ref.~\cite{MP1980}, the weak gauge-field dynamics
aligns the vacuum such that the corresponding gauge fields
remain massless, and the subgroup $SU(2)_\c\times SU(2)_\l$ is unbroken; no
dynamical Higgs mechanism is taking place.
This disproves recent claims in the literature \cite{CV2013}.

The inequality of Ref.~\cite{EW1983}, which guarantees
that $C>0$ in the broken phase, makes this a rigorous result.
We stress that the 8-flavor theory can be regularized
such that all the conditions of Ref.~\cite{EW1983} are fulfilled.
According to universality, the resulting effective potential, Eq.~(\ref{Vweak}),
must be independent of all details of the lattice regularization.

In order to keep this paper self-contained we have included a rederivation
of the most general order-$g^2$ continuum effective potential
in App.~\ref{Veffcont} (for a recent review, see Ref.~\cite{RC2010}).
As explained in the appendix,
when applying the master formula~(\ref{ChQCED})
we have to treat differently those generators that are proportional to $\g_5$
on the basis we are using, and those that are not.
In the appendix we illustrate this by working out explicitly
the case where an abelian gauge field is weakly coupled
to the generator $T_3^-$.  We show that, even though this generator
is axial with respect to the basis that diagonalizes the same-site
mass matrix (see Eq.~(\ref{PQa})), the true vacuum re-aligns itself
along the one-link mass term, and the abelian symmetry
ends up being vectorial and unbroken.

\subsection{\label{latt8} Comments on the lattice theory}
The result of the previous subsection
was derived after taking the continuum and chiral limits.
In a numerical lattice computation, there are usually practical considerations
dictating the use of non-vanishing mass terms.
In addition, discretization effects are unavoidable.

Let us momentarily turn off the weak gauge couplings $g_\c$ and $g_\l$ and thus the associated
continuum effective potential~(\ref{Vweak}), as well as any mass terms.
In other words, let us consider the chiral limit at non-zero lattice
spacing, with $SU(2)_\c\times SU(2)_\l$ a global symmetry group.
The order-$a^2$ staggered effective potential corresponding to this
situation is recorded in App.~\ref{Veffasq}.\footnote{%
  The symmetries of staggered fermions forbid order-$a$ terms in the
  effective potential.
}
It should be clear from the complicated form
of this effective potential that many orientations of the vacuum
will be inequivalent on the lattice, even if they become equivalent
in the massless continuum theory for $g_\c=g_\l=0$.
As discussed in App.~\ref{Veffasq},
one can easily envisage values for the order-$a^2$ low-energy constants
(LECs) that would prefer
the vacuum $\S_0$, and others that would prefer the vacuum $\S_1$.\footnote{%
  Note the change of basis of the $\S$ field performed in App.~\ref{Veffasq}.
}

We do not know the actual values of the LECs of the 8-flavor
staggered theory for a given number of colors.  There is a clear
message, however, that does not require this knowledge.  In general,
the vacuum of the theory will be influenced by all sources:
discretization effects, explicit mass terms, and weak gauge fields.
In the region where they are comparable,
\begin{equation}
  m/\L \sim a^2\L^2 \sim g^2_\c \sim g^2_\l \ ,
\label{params}
\end{equation}
one would expect a complicated phase diagram.
In the previous subsection we considered the limiting case where
\begin{equation}
  m/\L , a^2\L^2 \ll g^2_\c , g^2_\l \ .
\label{contparams}
\end{equation}
By contrast, the opposite limit
\begin{equation}
  m/\L , a^2\L^2 \gg g^2_\c , g^2_\l \ ,
\label{lattparams}
\end{equation}
will be dominated by discretization effects that in general will
have nothing to do with the continuum physics we are after.
According to one example we give in App.~\ref{Veffasq}, the discretization effects
prefer the $\S_0$ vacuum associated with the same-site mass term.
Close enough to the continuum limit, we would then expect a cross-over
from the vacuum $\S\sim\S_0$ to the true continuum vacuum $\S_1$
of the weakly gauged theory.
Thus, ensuring that a lattice study is conducted with the correct
parameter hierarchy~(\ref{contparams}), as opposed to the
hierarchy~(\ref{lattparams}),
may not be an easy task.

\section{\label{6 flavors} Six flavors}
In this section we consider a 6-flavor theory.
As before, the fermion fields reside in the fundamental representation
of a strongly interacting $SU(N_c)$ gauge group with $N_c \ge 3$,
that confines at a scale $\L$.
In the absence of additional weak gauge fields, the massless continuum theory
has an $SU(6)_L\times SU(6)_R$ flavor symmetry, which is spontaneously broken
to the diagonal, vectorial subgroup $SU(6)_V$.

The lattice theory is constructed using three reduced staggered fields,
with two fields $\c_i$, $i=1,2$, defined by the projection~(\ref{proj}),
and a single field $\l$ defined by the alternative projection~(\ref{opproj}).
The non-abelian flavor symmetry is therefore $SU(2)=SU(2)_\c$.
Notice that the same lattice theory can also be viewed as composed
of one standard staggered fermion, say $\c_2+\l$,
and a single reduced staggered fermion, $\c_1$.

Our interest in this particular discretization arises because
the model with $N_c=3$ was recently investigated numerically in Ref.~\cite{CV2013}.
As in the 8-flavor case we will first study two choices for the mass matrix,
while keeping the $SU(2)$ flavor symmetry global.  We will then study
the weak coupling of $SU(2)$ to an additional gauge field, again finding
that the vacuum aligns such that this symmetry is unbroken.

The 6 Dirac flavors of the continuum theory emerge from the lattice fields
according to
\begin{equation}
\label{6fcl}
\c_1\to\j_1,\ \j_2\ ,\quad\c_2\to\j_3,\ \j_4\ ,\quad\l\to\j_5,\ \j_6\ .
\end{equation}
We first choose one-link mass terms~(\ref{Sredmass})
for all reduced staggered fields.
On the continuum basis above the resulting mass term will be
proportional to the $2\times 2$ identity matrix for each reduced
staggered field.  However, we will now allow the Dirac fields originating
from $\c_1$ to have a different mass from the rest, namely,
\begin{equation}
\label{Mdiag}
\int d^4x\left(m'\left(\bj_1\j_1+\bj_2\j_2\right)
+m\left(\bj_3\j_3+\bj_4\j_4+\bj_5\j_5+\bj_6\j_6\right)\right)\ .
\end{equation}

Alternatively, we may introduce a single-site mass term~(\ref{ssmasseq})
for the standard staggered field $\c_2+\l$, and a one-link mass term only
for the remaining reduced staggered fermion $\c_1$.
In the continuum limit, the mass term is now
\begin{equation}
\label{6fclmass}
\int d^4x\left(m'\left(\bj_1\j_1+\bj_2\j_2\right)+m\left(\bj_5\j_3+\bj_6\j_4+\bj_3\j_5+\bj_4\j_6\right)\right)\ ,
\end{equation}
which corresponds to the mass matrix
\begin{equation}
  M_0 =
  m\ \left(\begin{array}{cc}
    \x & 0 \\
    0 & \t_1
  \end{array}\right)
  \otimes I_2
  =m\
  \left(\begin{array}{cccccc}
    \x & 0 & 0 & 0 & 0 & 0 \\
    0 & \x & 0 & 0 & 0 & 0 \\
    0 & 0 & 0 & 0 & 1 & 0 \\
    0 & 0 & 0 & 0 & 0 & 1 \\
    0 & 0 & 1 & 0 & 0 & 0 \\
    0 & 0 & 0 & 1 & 0 & 0
  \end{array}\right)\ .
\label{massmat6}
\end{equation}
In the block form in the middle, $\t_1$ is the first Pauli matrix,
the upper-left entry is $\x=m'/m$, and the off-diagonal entries
represent $1\times 2$ and $2\times 1$ blocks of zeroes.

The mass matrix~(\ref{massmat6})
can be rotated to a positive diagonal matrix by a non-anomalous
$SU(6)_L\times SU(6)_R$ basis transformation.
Using the vectorial $SU(2)$ transformation
\begin{equation}
  P =
  \left(\begin{array}{cc}
    1 & 0 \\
    0 & \frac{1}{\sqrt{2}}(I_2-i\t_2)
  \end{array}\right)
  \otimes I_2 \ ,
\label{P6}
\end{equation}
which rotates $\j_3$ and $\j_4$ into
$\j_5$ and $\j_6$, the mass matrix is first brought to the form
\begin{equation}
  P^\dagger M_0 P =
  m\ \left(\begin{array}{cc}
    \x & 0 \\
    0 & \t_3
  \end{array}\right)
  \otimes I_2
  = m\ \mbox{diag}(\x,\x,1,1,-1,-1) \ .
\label{M26}
\end{equation}
We then apply the non-anomalous chiral rotation
\begin{equation}
\label{Q6}
  Q = \mbox{diag}(1,1,1,1,i\g_5,-i\g_5)
\end{equation}
arriving, analogous to the 8-flavor case, at
\begin{equation}
   QP^\dagger M_0 P Q = m\ \mbox{diag}(\x,\x,1,1,1,1) \ .
\label{finalM6}
\end{equation}

\subsection{\label{flavor6} Global lattice flavor symmetry}
The lattice $SU(2)$ flavor symmetry that rotates $\c_1$ into $\c_2$ will,
in the continuum limit, rotate $\j_1$ into $\j_3$  and $\j_2$ into $\j_4$.
Using the basis introduced in Eq.~(\ref{6fcl}),
the $SU(2)$ generators are
\begin{equation}
  T_a =
  \left(\begin{array}{cc}
    \t_a & 0 \\
    0  & 0
  \end{array}\right)
  \otimes I_2 \ .
\label{su2}
\end{equation}
We see that relative to the continuum basis where the one-link mass term
takes the form~(\ref{Mdiag}),
the $SU(2)$ transformations are vectorial, and unbroken provided that $m'=m$.

In contrast, the mass term~(\ref{6fclmass}) softly breaks the $SU(2)$ symmetry.
We may recast the $SU(2)$ generators $T_a$ of Eq.~(\ref{su2}) on the basis in which
the mass matrix~(\ref{massmat6}) takes the form~(\ref{finalM6}),
obtaining
\begin{subequations}
\label{su2PQ}
\begin{eqnarray}
  T'_1 = Q^\dagger P^\dagger T_1 P Q
  &=&
  \frac{1}{\sqrt{2}}\left(\begin{array}{ccc}
   0 & I_2 & -i\g_5\t_3 \\
   I_2 & 0 & 0 \\
  i\g_5\t_3 & 0 & 0
  \end{array}\right) \ ,
\label{su2PQa}\\
  T'_2 = Q^\dagger P^\dagger T_2 P Q
  &=&
  \frac{1}{\sqrt{2}}\left(\begin{array}{ccc}
   0 & -iI_2 & -\g_5\t_3 \\
   iI_2 & 0 & 0 \\
  -\g_5\t_3 & 0 & 0
  \end{array}\right) \ ,
\label{su2PQb}\\
  T'_3 = Q^\dagger P^\dagger T_3 P Q
  &=&
  \half\left(\begin{array}{ccc}
   2I_2 & 0 & 0 \\
   0 & -I_2 & i\g_5\t_3 \\
   0 & -i\g_5\t_3 & -I_2
  \end{array}\right) \ .
\label{su2PQc}
\end{eqnarray}
\end{subequations}
We see that, relative to the ``canonical'' basis defined by
the mass matrix~(\ref{finalM6}),
the $SU(2)$ group of Eq.~(\ref{su2}) has turned into an admixture
of vectorial and axial transformations.

\subsection{\label{gauge6} Gauging $SU(2)$}
As was done in Ref.~\cite{CV2013}, one may promote the $SU(2)$
lattice flavor symmetry to a local symmetry by introducing a new gauge field
$V_{\m a}$, with a coupling $g$ that is weak at the confinement scale $\L$.
Integrating out the weak gauge field gives rise to the continuum
effective potential
\begin{equation}
\label{Vweak6}
V_{\rm weak}(\S)=-g^2C\,\sum_a\tr\left(\S T_a\S^\dagger T_a\right)\ ,
\end{equation}
with, now, $\S\in SU(6)$.  Again $C$ is a positive low-energy constant \cite{EW1983}.

Let us now compare the vacua $\S_0$ and $\S_1$
defined by taking the chiral limit
with the mass terms~(\ref{6fclmass}) and~(\ref{Mdiag}), respectively.
Explicitly, these vacua are
\begin{equation}
\label{candvac6}
\S_0=
  \left(\begin{array}{cc}
    1 & 0 \\
    0 & \t_1
  \end{array}\right)
  \otimes I_2\ ,\qquad
\S_1=I_6 \ .
\end{equation}
Notice that these vacua have a larger symmetry than the mass terms
from which they have emerged, because the chiral limit does not depend
on the ratio $\x=m'/m$.
We find that
\begin{subequations}
\label{Vweakvalues6}
\begin{eqnarray}
V_{\rm weak}(\S_0)&=&-2g^2C\ ,
\label{Vweakvalues6a}\\
V_{\rm weak}(\S_1)&=&-12g^2C\ .
\label{Vweakvalues6b}
\end{eqnarray}
\end{subequations}
The conclusion is analogous to the previous section.
The true vacuum is $\S_1$.  It is the orientation that was selected
by choosing one-link mass terms for all reduced staggered fields.
Once again the vacuum aligns such that the gauged $SU(2)$ flavor group
is unbroken.  In terms of the continuum theory, we have therefore
gauged a subgroup of the unbroken diagonal $SU(6)_V$
flavor symmetry group.

It follows that the apparent ``Higgsing'' of the weak gauge fields
claimed in Ref.~\cite{CV2013} must be a lattice artifact, caused by
contributions to the effective potential that vanish in the continuum
limit.

\section{\label{conclusion} Conclusion}
A strongly coupled theory with multiple standard or reduced staggered fermions has
a lattice flavor symmetry group which is smaller than the flavor
symmetry group of its continuum limit.   If all fermions are massless,
some of the lattice flavor symmetries will have generators of the form
$T_a^\e=T_a\e(x)$, where $T_a$ is an element of some Lie algebra, and
$\e(x)$ is defined in Eq.~(\ref{eps}).   Whether such a symmetry should
be interpreted as a vector or an axial symmetry in the continuum limit
depends on the mass terms that may be added to the theory, as explained
in Sec.~\ref{basics}.   If the massless limit is taken
after the continuum limit, the embedding of the flavor symmetry
of the lattice theory into the
larger flavor symmetry of the continuum theory will depend
on the mass terms originally chosen on the lattice.
Of course, in the continuum
limit this is irrelevant, because the flavor symmetry emerging in that
limit will always be the same.   In both concrete examples considered in this
article, the emerging symmetry is $SU(N_f)_L\times SU(N_f)_R$, with
$N_f=8$ or $N_f=6$, spontaneously broken to the diagonal subgroup $SU(N_f)$
in the massless limit.

The situation changes if one chooses to gauge the lattice flavor symmetry
group, or a subgroup of it.   With staggered fermions,
global symmetries with generators $T_a^\e$ may also be gauged.
Since it is customary to interpret these symmetries as axial
symmetries, this raises the intriguing prospect of obtaining an
exact chiral gauge group from the lattice.   Moreover, since the strong
dynamics spontaneously breaks axial symmetries in the massless
limit, naturally a Higgs mechanism would take place, with the
weak gauge fields coupled to the $T_a^\e$ acquiring a mass.\footnote{%
  Undoing this Higgs mechanism by simply turning off the strong
interactions would then lead to a genuine chiral gauge theory with
unbroken gauge symmetry on the lattice!}
Reference~\cite{CV2013} claims to find evidence for this mechanism
from numerical studies of the 6-flavor and 8-flavor theories we discussed
in this article.

However, the analysis of Ref.~\cite{MP1980} of the effective potential generated
by the weak gauge fields,
combined with the rigorous inequality of Ref.~\cite{EW1983}, implies that this
cannot happen in the continuum limit.
In making this statement we are, of course, invoking universality
in that we assume that the form of the continuum effective potential
must be independent of all details of the lattice regularization.

It is the dynamics of the weak gauge
fields themselves that gives rise to vacuum alignment.   The true vacuum
aligns such that all the lattice flavor symmetries
that have been weakly gauged, including those generated
by the $T_a^\e$, become unbroken vector symmetries in the continuum limit.
Indeed, as we have explained in detail, it is always possible to
choose mass terms for the staggered fields such that all fermions
will be massive while none of the lattice flavor symmetries
are broken by these mass terms.   It follows that none
of these symmetries will be spontaneously broken when these mass terms
are taken to zero.  In other
words, the flavor structure of the lattice theory will always make it
possible for ``complete'' vacuum alignment to take place, so that
all the gauge fields that were coupled weakly to lattice flavor currents
remain massless.

It follows that the numerical evidence presented in Ref.~\cite{CV2013}
must be the consequence of lattice artifacts.  Indeed, away from the
continuum limit,
lattice artifact contributions to the effective potential for the vacuum
may compete with the contribution generated by the dynamical
flavor gauge fields.   A more detailed study of the effective potential
along the lines of Ref.~\cite{AW2004} is possible, but outside the scope
of this article.

A competition between lattice artifacts and the dynamics of weak gauge fields
may arise for other fermion formulations as well.  For a study of these
effects with Wilson fermions, we refer to Ref.~\cite{GS2014}.

\vspace{3ex}
\noindent {\bf Acknowledgments}
\vspace{3ex}

We acknowledge discussions with Simon Catterall.
MG thanks the School of Physics and Astronomy of Tel Aviv University
and YS thanks the Department of Physics and Astronomy of San Francisco
State University for hospitality.
MG is supported in part by the US Department of Energy, and
YS is supported by the Israel Science Foundation under grants no.~423/09 and~449/13.

\appendix
\section{\label{Veffcont} The continuum effective potential}
In this appendix we rederive the continuum effective potential
for the non-linear $\S$ field induced by
a single weak gauge-boson exchange in the underlying theory.
This can be done via an elegant spurion trick \cite{SP2002}.\footnote{%
  For early discussions of the continuum effective potential,
  see for example Ref.~\cite{veff}.
}

As usual we will take the strong sector to be an $SU(N_c)$ gauge
theory with $N_c\ge 3$, coupled to $N_f$ Dirac fields
in the fundamental representation.\footnote{%
  The derivation in this appendix applies to any $N_f\ge2$.
}
The global symmetry of the massless theory
is $SU(N_f)_L\times SU(N_f)_R$.
We introduce global flavor spurions
$Q^L = Q^L_a T_a$, $Q^R = Q^R_a T_a$, where $a=1,2,\ldots,N_f^2-1$.
Under $g_{L,R}\in SU(N_f)_{L,R}$ they transform as
$Q^{L,R}\to g_{L,R} Q^{L,R} g_{L,R}^\dagger$.
The partition function is
\begin{equation}
  Z(Q^L,Q^R)
  = \int  d[A] d[W] d[\j] d[\bj] \,
  \exp[-S_{\rm S}(A_\m,\j_i,\bj_i)-S_{\rm W}(W_\m,\j_i,\bj_i,Q^L,Q^R)] \ ,
\label{ZQECD}
\end{equation}
where $S_{\rm S}$ is the action for the strong dynamics,
with $A_\m$ the $SU(N_c)$ gauge field, and $\j_i,\bj_i,$ $i=1,2,\ldots,N_f,$
the quark fields.  The weakly coupled dynamics
is accounted for by%
\footnote{The factor of $i$ on the first line of Eq.~(\ref{QweakS})
is erroneously missing in the published version.}
\begin{eqnarray}
  S_{\rm W} &=&  \frac{1}{4} (\partial_\m W_\n - \partial_\n W_\m)^2
  + i g W_\m (Q^L_a J^L_{\m a} + Q^R_a J^R_{\m a}) \,,
\label{QweakS}\\
  J^R_{\m a} &=& \half \bj_i \g_\m (1+\g_5) T_{aij} \j_j \,,
\nonumber\\
  J^L_{\m a} &=& \half \bj_i \g_\m (1-\g_5) T_{aij} \j_j \,.
\nonumber
\end{eqnarray}
The partition function $Z(Q^L,Q^R)$ is invariant under
global $SU(N_f)_L\times SU(N_f)_R$ transformations.  The flavor
indices are carried by the spurions $Q^L,Q^R$, while the $W_\m$ is a single
gauge field, inert under the flavor transformations.
We get away with not having a full set of flavored gauge
fields because we are only aiming to extract the effect of a single
weak gauge-boson exchange.

To order $g^2$, the most general effective potential consistent
with the flavor symmetry is
\begin{equation}
  V_{\rm eff}(\S) = g^2 C_{RR}\, \tr(Q^R Q^R) + g^2 C_{LL}\, \tr(Q^L Q^L)
        - g^2 C_{LR}\, \tr(Q^L \S Q^R \S^\dagger) \,.
\label{ChQCED}
\end{equation}
The only part that depends on the non-linear field is the last term.
The corresponding LEC, $C_{LR}$, may be isolated by assuming
that the vacuum state is the identity matrix $I_{N_f}$, so that\footnote{%
  Here we assume the standard orthogonality relation
$\tr(T_a T_b) = \half \d_{ab}$.  Notice that the generators discussed
in the main text are normalized differently.
}
\begin{equation}
  \frac{\partial}{\partial Q^L_a}
  \frac{\partial}{\partial Q^R_b}\, V_{\rm eff}(I_{N_f})
  = -\frac{g^2}{2} \d_{ab} C_{LR} \,.
\label{getCLR}
\end{equation}
In order to relate $C_{LR}$ to the microscopic theory we apply
the same differentiations to the partition function $Z(Q^L,Q^R)$, finding%
\footnote{The factor of 3 in Eq.~(\ref{CLRmic}) comes from tracing over
the transverse projector.  It is erroneously missing in the published version.}
\begin{equation}
  C_{LR} = \frac{3}{16\p^2} \int_0^\infty dq^2 q^2\, \P_{LR}(q^2) \,,
\label{CLRmic}
\end{equation}
where
($P^\perp_{\m\n}$ is the transverse projector)
\begin{equation}
  \half\d_{ab}\, q^2 P^\perp_{\m\n}\, \P_{LR}(q^2) \rule{0ex}{3ex}
  = -\int d^4x\, e^{iqx} \svev{J_{\m a}^L(x) J_{\n b}^R(0)} \ .
\label{formLR}
\end{equation}
According to Ref.~\cite{EW1983}, $\P_{LR}(q^2)\ge0$, and so is $C_{LR}\ge0$.

We next explain how to use the master formula~(\ref{ChQCED}) when
various subgroups of the flavor symmetry group are weakly gauged.
As a first example, let us weakly gauge the $SU(2)_\c$ of Sec.~\ref{8 flavors}.
We obtain the contributions of the weak gauge fields $V_{\m a}$, $a=1,2,3,$ to
the effective potential of Eq.~(\ref{Vweak}), one at a time, as follows.
On the continuum basis~(\ref{id}), the weak gauge field $V_{\m a}$ couples
to a vector current $J^L_{\m a}+J^R_{\m a}$, with a generator given explicitly
in Eq.~(\ref{su2sa}).  We therefore set $Q_a^L=Q_a^R=1$ for the left- and
right-spurions associated
with this particular generator, while setting to zero all other spurions.
With the obvious identification $g^2 C_{LR} \to g_\c^2 C$, after
summing over the 3 generators, we obtain the first term on the right-hand
side of Eq.~(\ref{Vweak}).  The same argument applies to the second term.

As another example, suppose that we weakly gauge only the $U(1)$ symmetry
generated by $T_3^-$ of Eq.~(\ref{avb}), with coupling constant $e$.
We will work out the vacuum energies for the two vacua $\S_{0,1}$
of Sec.~\ref{8 flavors}.  We first do the calculation using, as before,
the basis~(\ref{id}).  On this basis, the abelian gauge field couples to
a vector current whose associated generator is given explicitly
in Eq.~(\ref{avb}).  Following the same steps, the effective potential is
\begin{equation}
  V_{\rm weak} =
  -e^2 C_{LR}\,\tr\!\left(\S T^-_3\S^\dagger  T^-_3\right) \ , \qquad
  \mbox{one-link basis}\ .
\label{VU1}
\end{equation}
The vacuum energies are
\begin{subequations}
\label{VweakU1}
\begin{eqnarray}
V_{\rm weak}(\S_0) &=& +8 e^2C_{LR}\ ,
\label{VweakU1a}\\
V_{\rm weak}(\S_1) &=& -8 e^2C_{LR}\ .
\label{VweakU1b}
\end{eqnarray}
\end{subequations}
As expected, the vacuum aligns with $\S_1$,
so that the $U(1)$ symmetry is vectorial and unbroken.

Let us repeat the calculation, but now using the basis in which
the same-site mass term is diagonal, Eq.~(\ref{finalM}).  According to
Eq.~(\ref{PQa}), on this basis the generator $T^-_3$ is axial,
which implies that we now have $Q^L=-Q^R\equiv\tT^-_3$ in Eq.~(\ref{ChQCED}).
Therefore, this time we find
\begin{equation}
  V_{\rm weak} =
  +e^2 C_{LR}\,\tr\!\left(\S \tT^-_3\S^\dagger  \tT^-_3\right) \ , \qquad
  \mbox{same-site basis}\ .
\label{VU0}
\end{equation}
The actual value $\tT^-_3$ of the spurions can be read off from the flavor
matrix that multiplies $\g_5$ in Eq.~(\ref{PQa}) for $a=3$, leading to
\begin{equation}
  \tT^-_3 = \tau_2 \otimes I_4 \ .
\label{tT}
\end{equation}
We next re-evaluate $V_{\rm weak}$ on the two vacua.  Now we must
use the expressions for $\S_{0,1}$ appropriate for the basis~(\ref{finalM}).
The vacuum oriented along the same-site mass term is
$\S_0=I_8$, and plugging this into Eq.~(\ref{VU0}) reproduces Eq.~(\ref{VweakU1a}).
Analogous to Eq.~(\ref{finalM}),
the vacuum oriented along the one-link mass term is now
\begin{equation}
  \S_1 = QP^\dagger I_8 P Q = Q^2 = \tau_3 \otimes I_4 \ ,
\label{1link0basis}
\end{equation}
and plugging this into Eq.~(\ref{VU0}) reproduces Eq.~(\ref{VweakU1b}).

As it must be, the vacuum energies are independent of the basis
we choose.  This example demonstrates explicitly that, even if
the weakly gauged (abelian) generator looks axial on some basis,
the true vacuum will re-orient itself such that, relative to it,
that generator is vectorial and unbroken.

\section{\label{Veffasq} Staggered effective potential at order $a^2$}
When writing down the staggered low-energy effective theory it is customary
to use a basis for the $\S$ field in which
the same-site mass term is diagonal in flavor (or taste) space.
Applying the change of basis $\S\to QP^\dagger \S P Q$
to the non-linear field introduced in Sec.~\ref{gauge} (\seef\ Eq.~(\ref{finalM})),
the order-$a^2$ staggered effective potential for the 8-flavor theory is
\cite{LS1999,AB2003,AW2004}
\begin{equation}
  \cv = \cu + \cu\,' \ ,
\label{V}
\end{equation}
where
\begin{eqnarray}
  -\cu  & = &
  C_1 \tr\! \left(\xi^{(2)}_5\Sigma\xi^{(2)}_5\Sigma^{\dagger} \right) \nonumber \\
  & & +\frac{C_3}{2} \sum_{\nu}\left[ \tr (\xi^{(2)}_{\nu}\Sigma
    \xi^{(2)}_{\nu}\Sigma) + \hc\right] \nonumber \\
  & & +\frac{C_4}{2} \sum_{\nu}\left[ \tr(\xi^{(2)}_{\nu 5}\Sigma
    \xi^{(2)}_{5\nu}\Sigma) + \hc\right] \nonumber \\
  & & +\,C_6 \sum_{\mu<\nu} \tr(\xi^{(2)}_{\mu\nu}\Sigma
  \xi^{(2)}_{\nu\mu}\Sigma^{\dagger}) \, ,
\label{U}
\end{eqnarray}
\begin{eqnarray}
  -\cu\,'  & = & \frac{C_{2V}}{4}
  \sum_{\nu} \left[ \tr(\xi^{(2)}_{\nu}\Sigma)
    \tr(\xi^{(2)}_{\nu}\Sigma)  + \hc\right] \nonumber \\
  &&+\frac{C_{2A}}{4} \sum_{\nu} \left[ \tr(\xi^{(2)}_{\nu
      5}\Sigma)\tr(\xi^{(2)}_{5\nu}\Sigma)  + \hc\right] \nonumber \\
  & & +\frac{C_{5V}}{2} \sum_{\nu} \left[ \tr(\xi^{(2)}_{\nu}\Sigma)
    \tr(\xi^{(2)}_{\nu}\Sigma^{\dagger}) \right]\nonumber \\
  & & +\frac{C_{5A}}{2} \sum_{\nu} \left[ \tr(\xi^{(2)}_{\nu5}\Sigma)
    \tr(\xi^{(2)}_{5\nu}\Sigma^{\dagger}) \right]\ .
\label{Uprime}
\end{eqnarray}
Here
\begin{equation}
  \xi_B^{(2)} =
  \left( \begin{array}{cc}
    \xi_B & 0 \\
    0 & \xi_B
  \end{array} \right)\ ,
\label{xi8}
\end{equation}
and
\begin{equation}
  \{\xi_B \} = \{I,\xi_{\mu},\xi_{\mu<\nu},\xi_{\mu 5},\xi_5 \}\ ,
\label{xiB}
\end{equation}
is a basis for $4\times 4$ hermitian matrices in flavor space,
constructed in the usual way from the matrices $\xi_{\mu}$ satisfying
the Dirac algebra.

Depending on the actual values of the LECs, the vacuum state will
have different orientations.  A sufficient condition
that the vacuum be oriented along the same-site mass term
(in the basis used here, this is the identity matrix $I_8$),
is that $C_1,$ $C_3,$ $C_4$ and $C_6$ are all positive,
while $C_{2A,V}=C_{5A,V}=0$.  A different parameter range, where
the vacuum is oriented with the one-link mass term,
is when $C_{2V}$ and $C_{5V}$ are positive,
and the remaining LECs vanish.

\vspace{5ex}

\end{document}